\documentclass[aps,prb,preprint,groupedaddress]{revtex4-1}
\usepackage{graphicx}
\usepackage{textcomp}
\usepackage{gensymb}
\usepackage{amsmath}

\begin{document}
\title{Negative  nonlinear damping of a graphene mechanical resonator}

\author{Vibhor~Singh}
\author{Olga~Shevchuk}
\author{Ya.~M.~Blanter}
\author{Gary~A.~Steele}
\email{g.a.steele@tudelft.nl}
\affiliation{Department of Quantum Nanoscience, Kavli Institute of Nanoscience, Delft University of Technology, Lorentzweg 1, 2628 CJ Delft, The Netherlands}
\date{\today}

\begin{abstract}
We experimentally investigate the nonlinear response of a multilayer graphene resonator using a superconducting microwave cavity to detect its motion. The radiation pressure force is used to drive the mechanical resonator in an optomechanically induced transparency configuration. By varying the amplitudes of drive and probe tones, the mechanical resonator can be brought into a nonlinear limit. Using the calibration of the optomechanical coupling, we quantify the mechanical Duffing nonlinearity.
By increasing the drive force, we observe a decrease in the mechanical dissipation rate at large amplitudes, suggesting a negative nonlinear damping mechanism in the graphene resonator.
Increasing the optomechanical backaction, we observe a nonlinear regime not described by a Duffing response that includes new instabilities of the mechanical response.
\end{abstract}

\maketitle

The unique properties of graphene such as atomic thickness, low mass density, and high modulus of rigidity make it very attractive material for nanoscale electromechanical systems (NEMS) for several technological applications.
After the first demonstration of few layer thick graphene NEMS \cite{bunch_electromechanical_2007}, there has been an extensive studies on graphene nanoelectromechanical systems ranging from electromechanical resonators \cite{chen_performance_2009,singh_probing_2010}, oscillators \cite{chen_graphene_2013} and optomechanical systems aiming to probe the quantum regime of graphene motion \cite{singh_optomechanical_2014,weber_coupling_2014,song_graphene_2014,PhysRevApplied.3.024004}.
In this pursuit, large mechanical quality factors in graphene based NEMS have been demonstrated as well \cite{eichler_nonlinear_2011,singh_optomechanical_2014}.
Due to its atomic thickness, graphene based NEMS also exhibit rich nonlinearity such as onset of Duffing nonlinearity and nonlinear damping at realativly small mechanical amplitudes \cite{eichler_nonlinear_2011,song_stamp_2012}.
These properties further makes graphene an attractive candidate for developing optomechanical systems to reach the quantum regime of graphene motion \cite{voje_generating_2012}, to store microwave photons \cite{zhou_slowing_2013}, and could possibly be useful to understand dissipation in graphene NEMS for improved device performance \cite{imboden_dissipation_2014}.

The coupling between mechanical resonator and optical/superconducting microwave cavities has enabled the detection of mechanical motion with excellent sensitivities \cite{teufel_nanomechanical_2009,anetsberger_measuring_2010,wilson_measurement_2014}, offering an attractive platform to characterize the nonlinear response of mechanical resonators. 
In this Letter, we study non-linear dynamics of a multilayer graphene resonator by means of coupling it to a superconducting microwave cavity.
The graphene resonator is driven by injecting two microwave tones in the cavity, which are detuned by the mechanical resonant frequency leading to an oscillating radiation pressure force which drives the mechanical resonator. By changing the amplitude of these tones, we can independently control the driving force and dissipation due to the optomechanical backaction forces. We drive the mechanical resonator into the Duffing regime and characterize the nonlinearity. With increase in the driving force, we observe a reduction in linear dissipation rate, large hysteresis with sweep direction, and an instability in the mechanical amplitude.

Our device consists of  a multilayer graphene resonator coupled to a superconducting microwave cavity as studied previously \cite{singh_optomechanical_2014}.
Fig. 1a shows a scanning electron microscope image of a multilayer graphene resonator coupled to a superconducting microwave cavity. The multilayer graphene mechanical resonator is 10~nm thick and is suspended above a gate electrode of the microwave feedline by approximately 150~nm.
The superconducting cavity is in a quarter wavelength coplanar waveguide geometry fabricated with an alloy of molybdenum and rhenium ($T_c\approx$~9~K) on an intrinsic silicon substrate \cite{singh_molybdenum-rhenium_2014}. 
The measurements are performed in a dilution refrigerator under vacuum at 14~mK. 
The superconducting cavity has a resonance frequency of $\omega_c=2\pi\times5.90054$~GHz, with an internal dissipation rate $\kappa_i=2\pi\times54$~kHz and coupled to a feedline with an external coupling rate $\kappa_e=2\pi\times188$~kHz (coupling fraction $\eta=\frac{\kappa_e}{\kappa(=\kappa_e+\kappa_i)}=0.78$).
The graphene resonator forms a mechanically compliant capacitor to the microwave feedline as shown schematically in fig. 1b. Motion of graphene resonator modulates the capacitance and hence the cavity frequency. The graphene resonator has a resonance frequency of $\omega_m=2\pi\times36.233~$MHz. Using thermal noise, we calibrate the optomechanical coupling defined as $g_0=\frac{d\omega_c}{dx}x_{zpf}$, where $x_{zpf}$ are the quantum zero-point fluctuations of the mechanical resonator to be $2\pi\times0.83$~Hz \cite{singh_optomechanical_2014}, which also provides an absolute calibration of displacement amplitudes.

In order to probe mechanical response, we take advantage of the optomechanical coupling and sideband resolved limit ($\omega_m\gg\kappa$) in an optomechanically induced transparency (OMIT) setup. 
In OMIT setup, two microwave fields are injected inside the cavity.
A strong drive field $p_{drive}$ at lower mechanical sideband frequency $\omega_d=\omega_c-\omega_m$ and  a weak probe field $p_{probe}$ measures the cavity response by sweeping the probe tone in the vicinity of $\omega_c$.
When the detuning between drive and probe fields $\Omega=\omega_p-\omega_d$ matches $\omega_m$, the mechanical resonator experiences coherent radiation pressure force.
Coherent response of the mechanical resonator to the radiation pressure force up-scatter the drive field exactly at $\omega_p$ leading to an interference with the original probe field measuring the cavity response. This phenomenon is called optomechanically induced transparency (OMIT) \cite{agarwal_electromagnetically_2010,weis_optomechanically_2010} as described schematically in Fig.~2a.
It is worth pointing out that unlike heterodyne mixing schemes with low frequency RF drive, the radiation pressure force drive eliminates the need to apply a dc gate voltage.
Furthermore, while the strength of the probe tone allows to control the driving force on the mechanical resonator, independently the drive tone can be used to tune the dissipation in the mechanical resonator using the optomechanically  backaction.

For an overcoupled single port cavity, the interference between the probe field and upconverted field (drive being at lower motional sideband) leads to an absorption feature at low driving powers as shown schematically in Fig.~2(b). In the linear response limit, the resulting reflection coefficient of the cavity can be written as, 
$S_{11}(\omega)=1-\eta\kappa\frac{\chi_c}{1+g^2\chi_m\chi_c}$
where $\chi_{m}(\omega)=\frac{1}{-i(\Omega-\omega_m)+\gamma_m/2}$ 
is the susceptibility of the mechanical resonator, $\chi_c = \frac{1}{-i(\Omega-\omega_m)+\kappa/2}$ is the suspectibility of the cavity, $\gamma_m$ is the mechanical dissipation rate, $g=g_0\sqrt{n_d}$ is the many-photon optomechanical coupling strength, and $n_d$ is the number of the drive photons. 
In the limit $\kappa\gg2g\gg\gamma_m$, the measurement of the optomechanical induced absorption (OMIA) allows to directly probe the responsivity of the mechanical resonator giving its amplitude and dissipation rate, thus making it a sensitive technique.

For a red-sideband drive $\omega_d=\omega_c-\omega_m$, the minimum value of the reflection coefficient is given by much simplified expression
$\left|\frac{2\eta}{1+C}-1\right|$, where optomechanical cooperativity $C$ is defined as $C=\frac{4g_0^2n_d}{\gamma_m\kappa}$. In the limit of no optomechanical coupling ($C=0$), we recover $|2\eta-1|$ expression for minimum for a single port reflection cavity, which sets the base line of OMIA feature. The linewidth  of absorption feature is given by $(1+C)\gamma_m$, where the additional term $C\gamma_m$ originates from the backaction effects of drive photons and can be tuned by $n_d$.
Furthermore, amplitude of the mechanical resonator can also be cast into a convenient form, $x=x_{zpf}\left(\frac{C}{1+C}\right)\left(\frac{\kappa_e}{g}\right)\sqrt{n_p}$. It is instructive to see that for low cooperativity $(C<1)$, the mechanical amplitude can be tuned by both the probe and drive tone as $x\propto\sqrt{n_dn_p}$. On the other hand, in the limit $C>1$,  the mechanical amplitude is proportional to $\sqrt{\frac{n_p}{n_d}}$, suggesting that an increase in drive field leads to optomechanical damping and hence a reduction in the mechanical amplitude. An increase in the probe field, however, in both cases drives the mechanical resonator harder and yields larger amplitude.

In Fig 3, we probe the OMIA response in detail by varying the number of intracavity probe photons $n_p$, hence the driving force, while keeping the number of drive photons fixed at $n_d=2.5\times10^7$ and $1.0\times10^8$.
At low number of probe photons, the OMIA feature is determined by the linear response of the mechanical resonator. As $n_p$ is increased further, the nonlinearity in the OMIA response becomes evident with a stiffening of the mechanical resonator (positive shift in the resonance frequency) and the shark-fin like Duffing response accompanied by hysteresis with respect to frequency sweep-direction.

In addition to the clear Duffing response,  with the exception of the
bottom two curves, it can also be seen that the OMIA dip on the
non-linear regime becomes deeper. Qualitatively, the observation of a
deeper OMIA dip when $n_p$ is increased can be understood from a
reduction of the mechanical damping rate as the resonator is driven to larger amplitudes. Such a decreased mechanical damping rate would give a larger cooperativity and thus a deeper OMIA dip. In the last two curves, the cooperativity is continuing to increase, but the OMIA dip becomes less deep as the cavity has now crossed over to an effective undercoupled regime (see supplementary info of ref \cite{singh_optomechanical_2014} for more details).
In addition to the deeper OMIA dip that is suggestive of a decreased
mechanical damping at higher drive forces, Fig~3 also shows
additional features. Comparing panels (a) and (b), smaller and larger
$n_d$ respectively, the mechanical linewidth in panel (b) is
significantly larger. This is a consequence of increased
optomechanical damping, which also explains the absence of hysteresis
and shows only a transition to a Duffing response at higher powers \cite{lifshitz_nonlinear_2008}.
Finally, in panel (b), at the highest drive forces, we also observe an instability in the response in the form of a spike in reverse frequency sweep.

To gain quantitative insight into these observations, we perform numerical fits on the data shown in Fig.~3. The nonlinear response can be primarily captured by including a Duffing term $\alpha x^3$ in the restoring force of the mechanical resonator \cite{shevchuk_optomechanical_2015}. Following Ref~\cite{shevchuk_optomechanical_2015}, we perform numerical fits to extract the linear mechanical dissipation rate ($\gamma_m$), mechanical amplitude ($x_0$), and the Duffing parameter ($\alpha$) for any given probe and drive power. The gray curve in Fig~4(a) is the numerical fitted curve overlaid on top of the experimentally measured data allowing us to extract the Duffing parameter to be $\alpha=2.3\times10^{15}$~kgm$^{-2}$s$^{-2}$. Using the analytical expression for the onset of Duffing bifurcation point
$\omega_{up}=\omega_m+\frac{3}{8}\frac{\alpha}{m_{eff}\omega_m}(x_{up}^2)$, we get $\alpha=2.5\times10^{15}$~kgm$^{-2}$s$^{-2}$, which is close to the result we get by performing numerical fits. Fig.~4(b, c) plot the linear damping rate with the amplitude of the resonator extracted by performing numerical fits on datasets shown in Fig.~3.
At low amplitude, we observe mechanical damping rates $\gamma_m$ of $2\pi\times700$~Hz ($Q_m~=~51760$) for $n_d~=2.5\times10^7$, while for higher amplitude, the damping rate drops to $2\pi\times410$~Hz ($Q_m~=~88373$).
At large number of probe photons, the nonlinear dynamics of the OMIA becomes far more complex. The model with Duffing term in the restoring force still captures the response except the instability (sharp absorption feature in Fig.~3(b)) in the reverse frequency sweep.

The decrease in observed damping rate at higher amplitudes suggests the presence of a negative non-linear damping term $\mu$$x^2\dot{x}$ term in the equation of motion of the mechanical resonator \cite{shevchuk_optomechanical_2015}. 
As this negative nonlinear damping occurs also at low cooperativities, and as it is not seen in the theoretical calculations treating the optomechanical nonlinear response, we do not believe that it is an optomechanical effect, but instead intrinsic to the graphene resonator. There has been also observations of nonlinear damping in nanomechanical resonators \cite{zaitsev_nonlinear_2005} and carbon based resonators \cite{eichler_nonlinear_2011}. One possible source of negative non-linear damping is the saturation of two-level-systems coupled to the mechanical resonator \cite{PhysRevB.72.224101,PhysRevB.81.184112}. At low drive powers, these two-level systems can absorb energy from the mechanical resonator, increasing the mechanical damping rate. At higher powers, the two-level systems become saturated, and the damping rate goes down. Such an process was suggested as an explanation of power-dependent attenuation losses in glasses \cite{narayanamurti_tunneling_1970,anderson_anomalous_1972,arnold_nonlinear_1974}, and also was used to describe power-depended dielectric losses in superconducting electrical resonators \cite{gao_noise_2007}. For such a saturation result in nonlinear damping effects, the level 
spacing of the TLSs should be larger than the bath temperature. In order for TLSs to describe the negative nonlinear damping observed here, the coupling between the TLSs and the mechanical resonator would have to be non-resonant, mediated by strong higher order processes.

In conclusion, we examined the nonlinear dynamics of a graphene resonator coupled to a superconducting microwave cavity. In linear response limit, optomechanically-induced transparency measurements easily allows us to extract linear damping rate and peak amplitude. At moderate driving force when response becomes nonlinear, we perform numerical fits by including a Duffing term in the mechanical restoring force and find $\alpha=2.3\times10^{15}$~kgm$^{-2}$s$^{-2}$. Increasing the driving force further, the OMIA response becomes complex and it is no longer captured by the Duffing term. At these large amplitudes, higher order nonlinearities start becoming relevant and make the mechanical damping rate to appear low at larger amplitudes, where we observe a qualitatively new phenomena of negative nonlinear damping in a mechanical resonator.


\section*{acknowledgments}
The authors would like to thank Andres Castellanos-Gomez, Sal Bosman, and Ben Schneider for their help during device fabrication and low-temperature measurements. The work was supported by the Dutch science foundation NWO/FOM and EC-FET Graphene Flagship.

\newpage

\begin{figure}
\includegraphics[width=120mm]{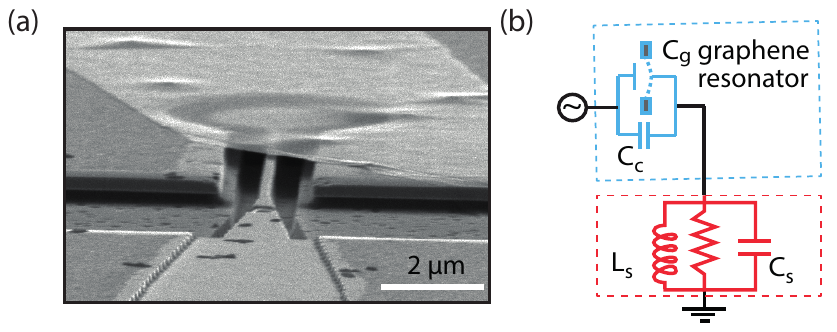}
\caption{(a) A scanning electron micrograph of a multilayer graphene (10 nm thick) drum-shape resonator coupled to a superconducting microwave cavity (not shown here). Graphene resonator is suspended 150~nm above the bottom gate electrode. (b) Schematic diagram of the device: graphene resonator couples external microwave radiation to the cavity by forming a coupling capacitor.}\label{fig1}
\end{figure}

\begin{figure}
\includegraphics[width=120mm]{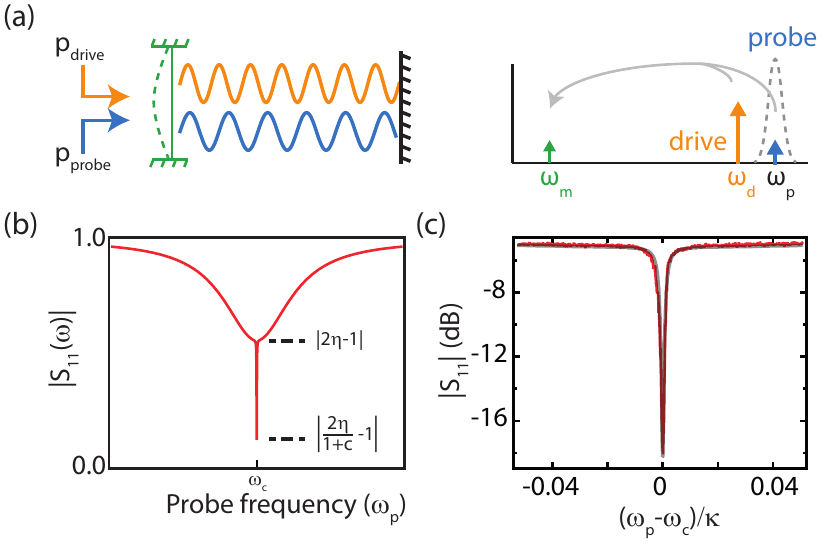}
\caption{(a) Schematic showing the idea of radiation pressure driving. Due to the optomechanical coupling, driving the cavity with a strong tone near $\omega_d=\omega_c-\omega_m$ and a weak probe tone $\omega_p$ near $\omega_c$ exerts a radiation pressure force on the mechanical resonator at $\omega_m$. The strength of radiation pressure force is controlled by the product of probe tone and the drive tone amplitudes. (b) Sketch of the cavity reflection coefficient in presence of a strong sideband drive. The optomechanical interaction produces an absorption feature in the cavity response. (c) A zoomed-in view of the OMIA feature showing the mechanical response in the linear regime.}\label{fig2}
\end{figure}

\begin{figure}
\includegraphics[width=85mm]{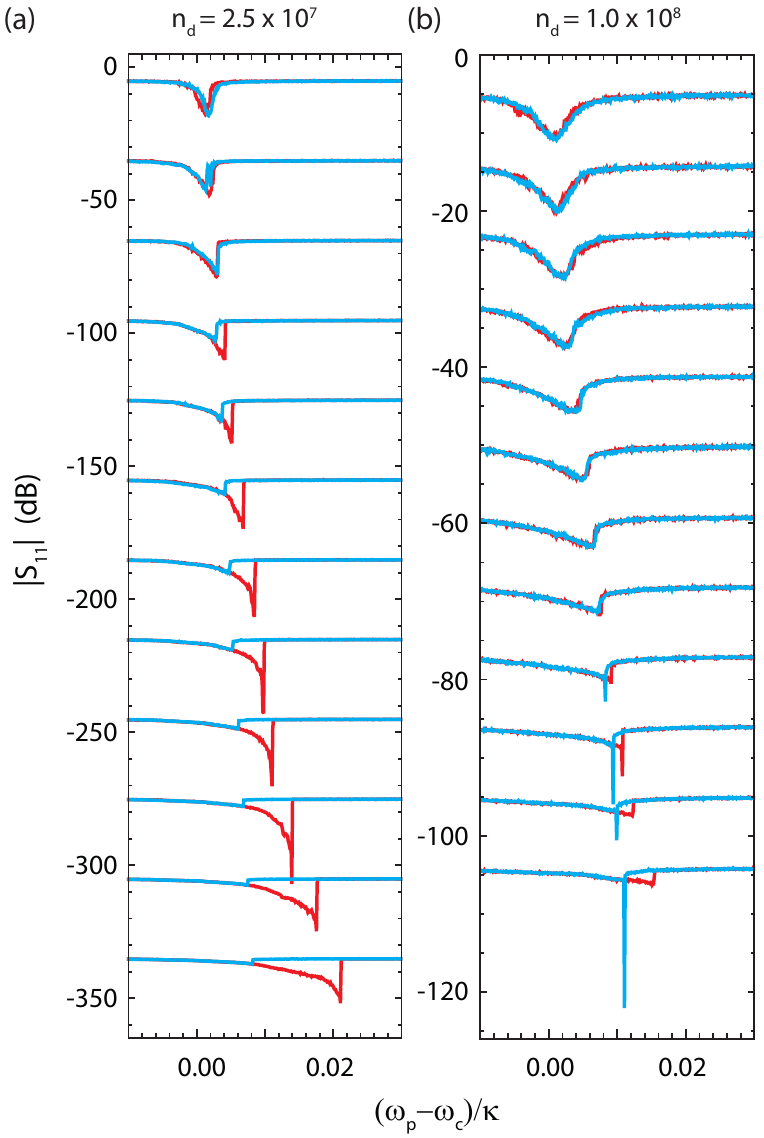}
\caption{Forward (red) and reverse (cyan) frequency sweep measurement of OMIA feature showing mechanical response at various probe and drive powers. The probe photons are swept from $n_p~=~2.5\times10^5$ to $3.14\times10^6$ in 1~dB steps (top to bottom). Number of drive photons $n_d$ is fixed at $2.5\times10^7$ for panel (a) and $1.0\times10^8$ photons for panel (b). The evolution of nonlinear response accompanied by the hysteresis can be clearly seen as probe power is increased (top to bottom). Panel (b) shows instability points as sharp dips appearing at large probe power. For clarity, measurements in (a) and (b) are plotted with offsets of -30~dB and -9~dB, respectively.}\label{fig3}
\end{figure}

\begin{figure}
\includegraphics[width=70mm]{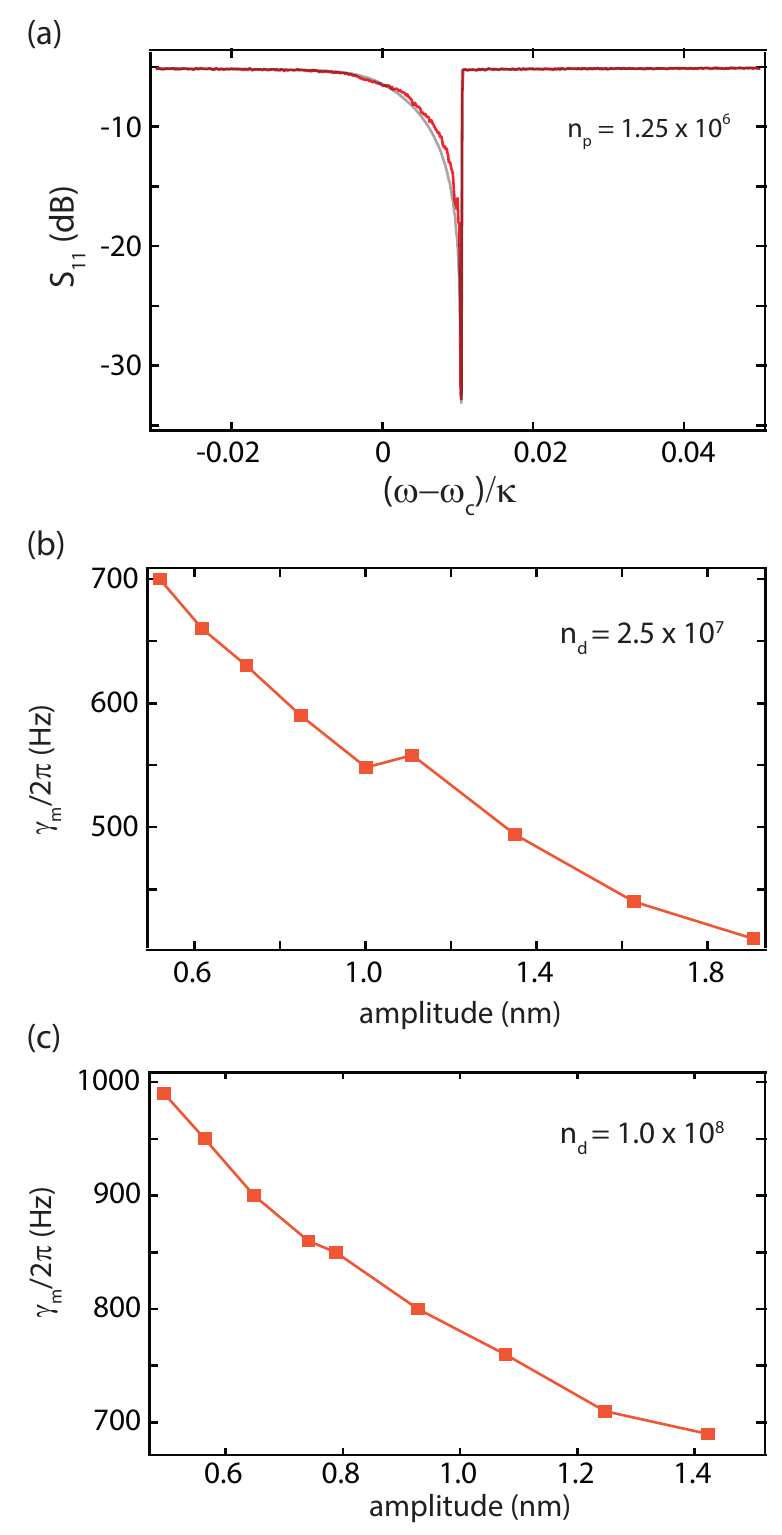}
\caption{(a) Measurement of $S_{11}$ showing strong nonlinear response (red curve) together with numerically fitted curve (gray) for $n_d=~=~2.5\times10^7$. (b, c) Extracted linear dissipation rate $\gamma_m$ plotted against mechanical amplitude for $n_d~=~2.5\times10^7$ and for $1.0\times10^8$, respectively. }\label{fig4}
\end{figure}

\end{document}